\documentclass[a4paper,11pt]{article}
\pdfoutput=1 

\usepackage{jcappub} 

\usepackage{aasmacros}
\usepackage[T1]{fontenc} 
\usepackage{cprotect}
\usepackage{multirow}
\usepackage{diagbox}
\usepackage{floatrow}

\newfloatcommand{capbtabbox}{table}[][\FBwidth]

\title{\boldmath Characterization of Extragalactic Point-Sources on E- and B-mode Maps of the CMB Polarization}

\author[a,b]{P. Diego-Palazuelos,}
\author[a]{P. Vielva,}
\author[a]{and D. Herranz}
\affiliation[a]{Instituto de F\'isica de Cantabria (CSIC-Universidad de Cantabria),\\ 
Avda. de los Castros s/n, E-39005 Santander, Spain}
\affiliation[b]{Dpto. de F\'isica Moderna, Universidad de Cantabria, \\
Avda. los Castros s/n, E-39005 Santander, Spain}

\emailAdd{diegop@ifca.unican.es}
\emailAdd{vielva@ifca.unican.es}
\emailAdd{herranz@ifca.unican.es}

\abstract{Although interesting in themselves, extragalactic sources emitting in the microwave range (mainly radio-loud active galactic nuclei and dusty galaxies) are also considered a contaminant from the point of view of Cosmic Microwave Background (CMB) experiments. These sources appear as unresolved point-like objects in CMB measurements because of the limited resolution of CMB experiments. Amongst other issues, point-like sources are known to obstruct the reconstruction of the lensing potential, and can hinder the detection of the Primordial Gravitational Wave Background for low values of $r$. Therefore, extragalactic point-source detection and subtraction is a fundamental part of the component separation process necessary to achieve some of the science goals set for the next generation of CMB experiments. As a previous step to their removal, in this work we present a new filter based on steerable wavelets that allows the characterization of the emission of these extragalactic sources. Instead of the usual approach of working in polarization maps of the Stokes' $Q$ and $U$ parameters, the proposed filter operates on E- and B-mode polarization maps. In this way, it benefits from the lower intensity that, both, the CMB, and the galactic foreground emission, present in B-modes to improve its performance. To demonstrate its potential, we have applied the filter to simulations of the future PICO satellite, and we predict that, for the regions of fainter galactic foreground emission in the $30$ GHz and $155$ GHz bands of PICO, our filter will be able to characterize sources down to a minimum polarization intensity of, respectively, $125$ pK and $14$ pK. Adopting a $\Pi=0.02$ polarization degree, these values correspond to $169$ mJy and $288$ mJy intensities.}

\begin{document}
\maketitle
\flushbottom

\section{Introduction}
\label{sec:introduction}

Although it is not their prime objective, Cosmic Microwave Background (CMB) experiments can also provide valuable information about the population of extragalactic sources that lies in the 20-800 GHz frequency range \citep[e.g., ][]{Core_ps}. Experiments at those frequencies open an observation window to the synchrotron emission coming from the relativistic jets of radio-loud active galactic nuclei and to dusty galaxies with a high star formation rate. In particular, the information contained in polarization allows the study of the strong magnetic fields present in both kinds of sources. Whereas the polarization degree of radio sources is well characterized at low frequencies, its nature is still poorly constrained at higher frequencies \citep{Galluzzi}, and, in general, little is known about the polarization degree of dusty galaxies due to the complex structure of galactic magnetic fields. Therefore, our understanding of the physics of these sources will greatly benefit from the plethora of experiments centered on the CMB polarization proposed for the next generation, like the CMB Stage-IV \citep{S4} or the PICO satellite \citep{PICO}.\\

However, from the point of view of a cosmologist, extragalactic sources are just an additional contaminant obscuring the signal of the CMB. Because of the limited resolution of CMB experiments (of the order of arcminutes), extragalactic sources appear as unresolved point-like objects in CMB maps, that, when uniformly distributed across the sky, behave like an additional white Gaussian noise at the angular power spectrum level. In this way, they can potentially become an important contaminant at small angular scales for frequencies up to $\sim 200$ GHz \citep{B-mode_detectability_Tucci,Puglisi, Trombetti}. At higher frequencies, and for low flux densities, dusty galaxies tend to cluster together, which introduces additional correlations into their angular power spectrum. Therefore, in addition to the removal of diffuse galactic foreground emission \citep{Dickinson,errard}, and the delensing of the secondary B-modes induced by weak gravitational lensing \citep{lensing_Lewis_Challinor, MyLensingPaper}, extragalactic point-source detection and subtraction is also a fundamental part of the component separation process necessary to achieve the science goals set for the next generation of CMB experiments. In particular, extragalactic point-sources would significantly affect the reconstruction of the lensing potential \citep{pST}, and consequently, severely limit the delensing of secondary B-modes.\\

The lensing potential, which is the projection onto the sphere of the integrated mass distribution along the line-of-sight between us and the last scattering surface \citep[e.g., ][]{lensing_Lewis_Challinor}, is an excellent probe for the matter distribution in the universe since it goes up to much higher redshifts than conventional galaxy surveys. Amongst other applications (e.g., see science goals pursued by \cite{simons_observatory}, \cite{S4}, \cite{PICO} or \cite{core_lensing}), a faithful estimate of the lensing potential could provide a measurement of the absolute mass scale of neutrinos \citep{planck2018lensing, neutrinos_and_cosmology}, and would help calibrate cluster masses to improve the interpretation of galaxy cluster surveys \citep{cluster_masses_w/_lensing, hu_cluster_masses, cluster_masses_statistical_approach, cluster_masses_measurements}. Although it can also be reconstructed from other large-scale structure tracers \citep{manzotti, multitracer} (like galaxy surveys \citep{delensing_w/_ska}, the Cosmic Infrared Background \citep{delensing_w/_cib, first_delensing_w/_CIB, delensing_sptpol_w/_herschel, planck2018lensing}, or tomographic line intensity mapping \citep{delensing_w/_line-intensity_mapping}), for the next generation of experiments, the best lensing potential reconstructions are expected to come from CMB data \citep{manzotti}.\\

When recovering the lensing potential through CMB measurements, there are at least two ways in which point-sources would affect the reconstruction. On the one hand, whether we use quadratic estimators \citep[e.g., ][]{hirata_seljak,qe} or \textit{maximum a posteriori} reconstructions \citep[e.g., ][]{lensit,lenseflow}, the small angular scales of polarization fields (and especially the ones from the EB cross-correlation) are the scales that contribute the most to the estimation of the lensing potential. As was previously mentioned, these are precisely the scales where, if not mitigated, point-source emission would dominate over the CMB. On the other hand, and given that point-sources are themselves tracers of the large-scale structure of the universe, they are also known to introduce spurious correlations between CMB fields and the lensing potential. Such correlations are further enhanced by the actual lensing of the point-sources' emission \citep{lensed_foregrounds}, because, as they come from cosmological distances, their emission is itself lensed by the rest of the matter distribution between them and us. Therefore, if they are not properly controlled, point-sources could lead to a poor and biased lensing potential reconstruction. Since they limit our ability to correctly estimate the lensing potential, point-sources would also condition the delensing of the secondary B-modes originated by weak gravitational lensing \citep{lensing_Lewis_Challinor, MyLensingPaper}.\\

All in all, point-sources could become an important obstacle for the detection of the Primordial Gravitational Wave Background (PGWB) for low values of $r$ \citep{B-mode_detectability_Tucci,Puglisi, Trombetti} due to, both, the noise-like signal they constitute in themselves, and the reduction in delensing power they cause by degrading lensing potential reconstructions. The detection of such PWGB, a relic background of stochastic gravitational waves that most inflationary models predict must have been produced during inflation (see e.g. \cite{Caprini} for a recent review), is one of the main science goals pursued by the next generation of CMB experiments. Well physically motivated inflationary models \citep[e.g.,][]{grishchuk, starobinski} predict that the amplitude of the signal that the PGWB leaves on the B-mode polarization of the CMB \citep{Polnarev, cabella_kamionkowski, Zhao}, which is controlled by the ratio between tensor and scalar perturbations $r=\mathcal{P}_t(k)/\mathcal{P}_s(k)$, should be of about $r\sim 0.001$. Current constrains put an upper limit to this value of $r<0.056$ \citep{planck_2018_inflation}. Unfortunately, a PGWB B-mode signal of such amplitude would fall far below the emissions of diffuse galactic foregrounds and extragalactic sources, and the secondary lensed B-modes. In this way, as we previously discussed, the improvement of the lensing potential estimates available for delensing and of component separation techniques are fundamental to achieve the goal of PGWB detection.\\

As a previous step to their removal, in this paper we present an alternative methodology to the usual approach of working in $Q$ and $U$ polarization maps \citep[e.g.,][]{argueso, PCCS_2013, PCCS_2015} by designing a filter based on steerable wavelets that is capable of characterizing extragalactic sources on E- and B-mode polarization maps. By working on maps of the B-mode polarization, we hope to take advantage of the lower intensity  that, both the CMB, and the galactic foreground emission, present in that channel in comparison to that of E-modes \citep{PolarizedDustForegrounds}, or $Q$ and $U$ maps. Although it could be repurposed for it, in principle, the filter has not been designed for blind source detection but rather for its application in the estimation of the polarization angle and intensity of already known point-sources. Working with already known sources is not a real limitation since, due to their low polarization degree ($\Pi\in[0.02,0.10]$), point-sources are expected to be at least $\sim 10-50$ times brighter in total intensity than in polarization ($\Pi=P/I$). Therefore, any source that is near the detection threshold in polarization intensity should be bright, and clearly detectable, in total intensity. In this way, we can always assume that sources were already identified in intensity beforehand. This is a common strategy in the study of the polarimetric properties of extragalactic sources (see for example \cite{lopez-caniego2009,PCCS_2015}). To demonstrate the potential of this new methodology, we have applied the filter to simulations of the future PICO satellite.\\

The work is structured as follows. In section \ref{sec:filter_design}, after showing the profile point-sources present on E- and B-mode maps, we specify the filter design, and describe the methodology devised for parameter estimation. The characterization of the methodology performance on simulations of the proposed PICO satellite is left for section \ref{sec:test_on_simulations}. Finally, the conclusions that can be drawn from this work, and some possible lines of future work, are discussed in section \ref{sec:conclusions}.\\

\section{Filter design}
\label{sec:filter_design}
    
In this section we will explain the details of the filter design, and present the methodology that will allow us to characterize extragalactic sources. We will start by introducing the mathematical expressions describing the profile of point-sources in E- and B-mode polarization maps in subsection \ref{sec:source_profile_in_polarization_maps}. Inspired by those profiles, we will then build a basis of steerable wavelet functions in subsection \ref{sec:wavelet_definition}, and show a simple method to recover from them the polarization angle and intensity of the source. Finally, we will acknowledge how working with discrete images affects the filter implementation in subsection \ref{sec:calibration}.\\

\subsection{Point-source profile in E- and B-mode polarization maps}
\label{sec:source_profile_in_polarization_maps}

As happens with the rest of galactic foregrounds, the emission of extragalactic sources is linearly polarized \cite[e.g.,][]{Dickinson, Tucci}, thus being fully characterized by its polarization angle $\phi$ and intensity $P$. In addition, because of the limited resolution of telescopes in the microwave range (of the order of arcminutes), extragalactic sources appear as point-like objects rather than extended structures since they cannot be resolved. Therefore, in maps of the $Q$ and $U$ Stokes' parameters, an extragalactic source located at an $\vec{r_i}$ position could be described like:
\begin{align}\label{eq:ideal point-like source}
Q(\vec{r}) = & \rho(\vec{r})P\cos2\phi, \notag\\
U(\vec{r}) = & \rho(\vec{r})P\sin2\phi,
\end{align}
with a radial profile $\rho(\vec{r})=\delta(\vec{r}-\vec{r}_i)$, and polarization angle defined between $\phi\in[0,\pi)$. Like it is customary for the study of compact sources in full-sky maps  \cite[e.g., ][]{argueso, PCCS_2013, PCCS_2015}, the filter will be applied to the projection onto the plane of a small square region of the sky containing the source. Working within this reduced surface allows for a better statistical characterization of the background surrounding the source, thus improving the performance of the filter. In addition, the small size and compact nature of point-sources ensure that, when pixels in the plane and in the sphere have approximately the same size, the projection will not introduce a significant distortion to the sources' shape.\\

In a first approximation, CMB experiments can be effectively modeled to have a circular Gaussian beam\footnote{Note that, in reality, CMB beams are generally elliptical instead of circular, and can have more complex profiles than the ones assumed here. Nevertheless, since the wavelets we are going to construct in section \ref{sec:wavelet_definition} are based on the profile point-sources have in polarization maps, they can be recalculated and adapted for any given beam.}. Because of this instrumental response, point-sources adopt a Gaussian profile characterized by the beam's \emph{Full Width at Half Maximum}, or alternatively, its $\sigma$ (related by $\mathrm{FWHM}=2\sqrt{2\ln2} \hspace{2pt} \sigma$):
\begin{equation}\label{eq:gaussian point-like source}
\rho(\vec{r})=\frac{1}{2\pi\sigma^2}e^{-r^2/2\sigma^2}.
\end{equation}
In this last equation, the coordinate origin was moved to the center of the source.\\

Following for instance \cite{cabella_kamionkowski}, when working in the plane, the source profile will transform from a pair of $Q$ and $U$ maps to the corresponding E- and B-mode maps like:
\begin{align}\label{eq:generic P_E P_B fourier}
\tilde{E}(\vec{q}) = &\frac{1}{2}[\cos2\theta \tilde{Q}(\vec{q})+\sin2\theta \tilde{U}(\vec{q})],  \notag\\
\tilde{B}(\vec{q}) = &\frac{1}{2}[\sin2\theta \tilde{Q}(\vec{q})-\cos2\theta \tilde{U}(\vec{q})],
\end{align}
where $\tilde{f}(\vec{q})$ stands for the Fourier Transform of a $f(\vec{r})$ function, and $\vec{q}=(q,\theta)$ are the polar coordinates in reciprocal space. Hence, introducing the Fourier Transforms of the discussed $Q(\vec{r})$ and $U(\vec{r})$ profiles in (\ref{eq:generic P_E P_B fourier}), the E- and B-mode profiles of the source in reciprocal space would be:
\begin{align} \label{eq:P_E P_B fourier}
\tilde{E}(\vec{q}) =&\frac{P}{2}[\cos2\theta \cos2\phi+\sin2\theta \sin2\phi]e^{-q^2\sigma^2/2}, \notag \\
\tilde{B}(\vec{q}) =&\frac{P}{2}[\sin2\theta \cos2\phi-\cos2\theta \sin2\phi]e^{-q^2\sigma^2/2}.
\end{align}
Calculating the inverse Fourier Transform of the previous expression results in a real space profile of:
\begin{align}\label{eq:P_E P_B real}
E(\vec{r}) =&\frac{P}{\pi}[\cos2\xi \cos2\phi +\sin2\xi \sin2\phi] \tau(r), \notag\\
B(\vec{r}) =&\frac{P}{\pi}[\sin2\xi \cos2\phi -\cos2\xi \sin2\phi] \tau(r),
\end{align}
where $\vec{r}=(r,\xi)$ are the polar coordinates in real space, and the radial dependence reads
\begin{equation}\label{eq:P_E and P_B radial dependence}
\tau(r)=\frac{1}{r^2}\left[e^{-r^2/2\sigma^2}\left(1+\frac{r^2}{2\sigma^2}\right)-1\right]. 
\end{equation}
To account for the different convention in the definition of E- and B-modes adopted by \cite{cabella_kamionkowski} in the plane, and by \verb|HEALPIX| \citep{healpix} in the sphere, the equations in (\ref{eq:P_E P_B real}) carry an extra 2 factor\footnote{If the reader were to repeat these calculations, they would obtain expressions like the ones shown in (\ref{eq:P_E P_B real}) but with a denominator of $2\pi$.} to make this formulation suitable for its application in real and simulated maps of the microwave sky.\\

\begin{figure}[tbp]
    \begin{center}
        \includegraphics[width=0.7\textwidth]{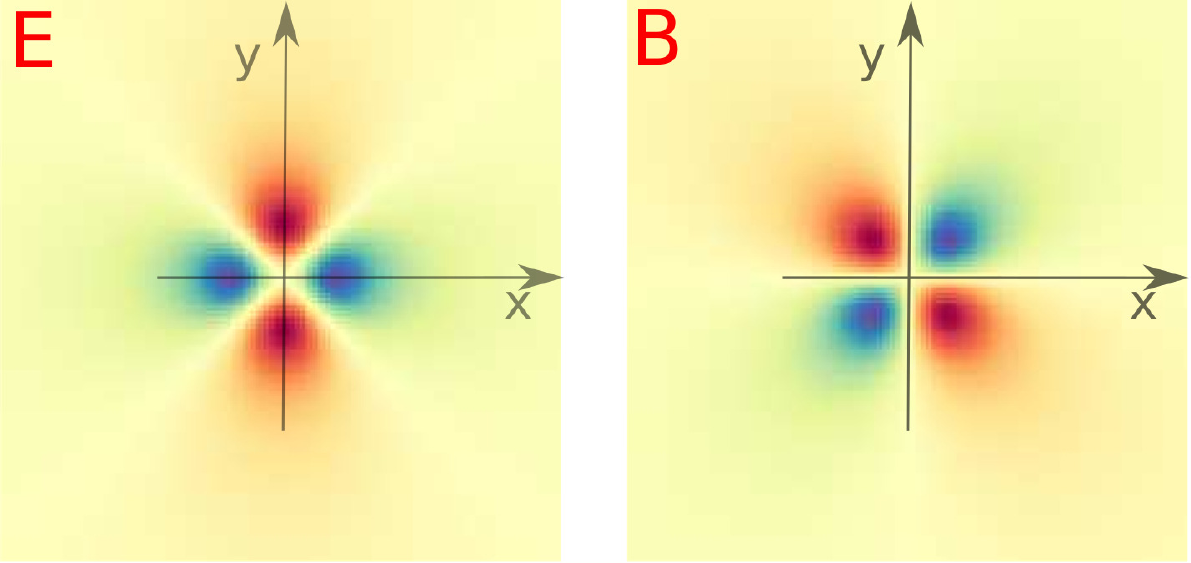}        
        \caption{\footnotesize $E(\vec{r})$ (left) and $B(\vec{r})$ (right) point-source profiles for a polarization angle of $\phi=0$.}
        \label{SrcProf}
    \end{center}
\end{figure}

As can be seen in figure \ref{SrcProf}, point-sources present a hot and cold two-lobes profile in E- and B-mode maps. The position of the hot and cold lobes is the opposite of what would be expected by just looking at the angular component of equations (\ref{eq:P_E P_B real}) because of the negative amplitude of its radial component $\tau (r)$. The symmetry between the sine and cosine terms in these equations, both for the polar ($\xi$) and polarization angles, introduces $\pi/4$ rotation relationships between the E- and B-mode profiles. Fixing the polarization angle, a $\pi/4$ spatial rotation transforms the E-mode profile into the B-mode one, $E(r,\xi\pm \pi/4,\phi)=\mp B(r,\xi,\phi)$. This property manifests itself in the plots shown in figure \ref{SrcProf}, and could be useful in cross-matching mechanisms between E- and B-modes to verify detections. Another useful relationship is $E(r,\xi,\phi\pm\pi/4)=\pm B(r,\xi,\phi)$, the equality between E- and B-modes under a $\pi/4$ rotation in the polarization angle.\\

These angular symmetries make $E(\vec{r})$ and $B(\vec{r})$ \textit{steerable functions}, i.e., functions that can be written as linear combinations of rotated versions of themselves (see \cite{freeman_adelson} for more insight about steerability conditions). Choosing as basis the $E(\vec{r})$ source profile for polarization angles 0 and $\pi/4$, 
\begin{align}\label{eq:P_x and P_y basis}
P_x(\vec{r}) =& E(\vec{r},\phi=0) = \frac{P}{\pi}\cos2\xi \tau(r), \notag\\
P_y(\vec{r}) =& E(\vec{r},\phi=\pi/4) = \frac{P}{\pi}\sin2\xi \tau(r),
\end{align}
it is immediate to see that indeed the source profile for any other polarization angle is just a rotation of this basis:
\begin{align}\label{eq:P_x and P_y combination E and B}
E(\vec{r})=&\cos2\phi P_x(\vec{r})+\sin2\phi P_y(\vec{r}), \notag\\
B(\vec{r})=&\cos2\phi P_y(\vec{r})-\sin2\phi P_x(\vec{r}).
\end{align}
The $x$ and $y$ nomenclature for basis functions was chosen to reflect along which axis do the cold lobes of the source fall. Here we have defined the basis functions starting from the E-mode profile of the source, but thanks to their angular symmetries, the very same basis could have been obtained from the B-mode profile: $P_x(\vec{r})= B(\vec{r},\phi=3\pi/4)$ and $P_y(\vec{r})= B(\vec{r},\phi=0)$.\\

\subsection{Wavelet definition and parameter estimation}
\label{sec:wavelet_definition}
    
Looking at $E(\vec{r})$ and $B(\vec{r})$ as written in equations (\ref{eq:P_x and P_y combination E and B}), one could intuitively think that a filter relying on the $P_x(\vec{r})$ and $P_y(\vec{r})$ functions may be the simplest approach to recover the polarization angle and intensity of the source. Going back to the definition of $\tilde{E}(\vec{q})$ in equations (\ref{eq:P_E P_B fourier}), imposing the 0 and $\pi/4$ polarization angles previously used to obtain the $P_x(\vec{r})$ and $P_y(\vec{r})$ basis, and computing the inverse Fourier Transform, leads us to a basis of (properly normalized) real space filtering functions:
\begin{align}\label{eq:filtering functions real space}
\psi_x(\vec{r},R)=&\frac{1}{2\pi R^2}\cos2\xi e^{-r^2/2R^2}, \notag\\
\psi_y(\vec{r},R)=&\frac{1}{2\pi R^2}\sin2\xi e^{-r^2/2R^2}.
\end{align}
The same basis of filtering functions could have been reached from the $\tilde{B}(\vec{q})$ profile in equations (\ref{eq:P_E P_B fourier}) if the $3\pi/4$ and $0$ polarization angles were chosen instead. These $\psi_x$ and $\psi_y$ filtering functions are just the 0 and $\pi/4$ rotations of the mother wavelet:
\begin{equation}\label{eq:mother wavelet real space}
\Psi(\vec{r},R)=\frac{1}{2\pi R^2}\cos2\xi e^{-r^2/2R^2}.
\end{equation}
We call this function a \textit{wavelet} not only due to the introduction of the scale $R$, but also because it is a compensated function:
\begin{equation}\label{eq:compensated wavelet}
\int_0^\infty \int_{0}^{2\pi} \Psi(\vec{r},R)rdrd\xi=\int_0^\infty re^{-r^2/2R^2}dr\int_{0}^{2\pi} \cos2\xi d\xi=0.
\end{equation}
Following \cite{wiaux}, compensated functions satisfy the admissibility condition, which in turn ensures the fulfilment of the synthesis condition. Therefore, $\Psi (\vec{r},R)$  fulfills all the conditions required to be a wavelet in the plane. In this way, we can study the source profile in E- and B-modes in terms of its decomposition into the wavelet coefficients of our two basis functions ($i=x,y$):
\begin{align}
\omega_i^E(\vec{r},R)=\iint E(\vec{r}')\psi_i(\vec{r}'-\vec{r},R) d\vec{r}', \notag \\
\omega_i^B(\vec{r},R)=\iint B(\vec{r}')\psi_i(\vec{r}'-\vec{r},R) d\vec{r}'.
\end{align}

Since we have defined $\psi_x(\vec{r},R)$ and $\psi_y(\vec{r},R)$ to have the same functional form than $P_x(\vec{r})$ and $P_y(\vec{r})$, we count with a steerable wavelet that will make possible the reconstruction of the source profile for any polarization angle through the linear combination of wavelet coefficients:
\begin{align}\label{eq:combination wavelet coefficients E and B}
\omega_E(\vec{r},\hat{\phi},R)=&\cos2\hat{\phi} \, \omega_x^E(\vec{r},R) +\sin2\hat{\phi} \, \omega_y^E(\vec{r},R), \notag\\
\omega_B(\vec{r},\hat{\phi},R)=&\cos2\hat{\phi} \, \omega_y^B(\vec{r},R) -\sin2\hat{\phi} \, \omega_x^B(\vec{r},R).
\end{align}
Therefore, the only remaining step of the filtering process would be to find a way to estimate a $\hat{\phi}$ value for the polarization angle.\\

The optimal method we found to estimate both, polarization angle and intensity, relies on the relationships the wavelet coefficients' central point keeps with these magnitudes. For the E-mode source profile, the $\omega_x^E(\vec{r},R)$ and $\omega_y^E(\vec{r},R)$ wavelet coefficients are:
\begin{align}\label{eq:psiEx and psiEy}
\omega_x^E(\vec{r},R) =&\frac{P}{8\pi^2}\frac{R^2}{\sigma^2+R^2}\Big[\cos2\phi e^{-z}+\Big(\cos4\xi \cos2\phi+\sin4\xi \sin2\phi\Big)\lambda(z,R)\Big], \notag \\
\omega_y^E(\vec{r},R)=&\frac{P}{8\pi^2}\frac{R^2}{\sigma^2+R^2}\Big[\sin2\phi e^{-z}+\Big(\sin4\xi \cos2\phi-\cos4\xi \sin2\phi\Big)\lambda(z,R)\Big],
\end{align} 
where the radial dependence reads
\begin{equation}\label{eq:radial dependence wavelet coefficients}
\lambda(z,R)=\frac{1}{2z^2}\Big[e^{-z}\Big(z(z+4)+6\Big)+2(z-3)\Big], \hspace{10mm}
 z=\frac{r^2}{2(\sigma^2+R^2)}.
\end{equation}
For both coefficients, if we focus our attention in the center of the image (when $r\rightarrow 0$, the radial terms tend to $\lambda(z,R)\rightarrow 0$ and $e^{-z}\rightarrow 1$), we are left with:
\begin{align}\label{eq:central pix psiEx and psiEy}
\omega_x^E(\vec{0},R)=&\frac{P}{8\pi^2}\frac{R^2}{\sigma^2+R^2}\cos2\phi, \notag\\
\omega_y^E(\vec{0},R)=&\frac{P}{8\pi^2}\frac{R^2}{\sigma^2+R^2}\sin2\phi.
\end{align}
Therefore, an estimation of the polarization angle can easily be computed through the ratio of central wavelet coefficients like:
\begin{equation}\label{eq:phi est E}
\hat{\phi}^E(R)=\frac{1}{2} \arctan\left(\frac{\omega_y^E(\vec{0},R)}{\omega_x^E(\vec{0},R)}\right).
\end{equation}
As it would be expected, the wavelet coefficients for B-modes are just a $\pi/4$ rotation of $\omega_x^E(\vec{r},R)$ and $\omega_y^E(\vec{r},R)$:
\begin{align}\label{eq:psiBx and psiBy}
\omega_x^B(\vec{r},R)=&\frac{P}{8\pi^2}\frac{R^2}{\sigma^2+R^2}\Big[-\sin2\phi e^{-z}+\Big(\sin4\xi \cos2\phi -\cos4\xi \sin2\phi \Big)\lambda(z,R)\Big], \notag\\
\omega_y^B(\vec{r},R)=&\frac{P}{8\pi^2}\frac{R^2}{\sigma^2+R^2}\Big[\cos2\phi e^{-z}-\Big(\cos4\xi \cos2\phi +\sin4\xi \sin2\phi \Big)\lambda(z,R)\Big].
\end{align}
Hence, an estimate of the polarization angle can be obtained from B-modes like:
\begin{equation}\label{eq:phi est B}
\hat{\phi}^B(R)=\frac{1}{2} \arctan\left(\frac{-\omega_x^B(\vec{0},R)}{\omega_y^B(\vec{0},R)}\right).
\end{equation}

Now that we count with the $\hat{\phi}^{E,B}$ estimates of the polarization angle, we can go back to the equations (\ref{eq:combination wavelet coefficients E and B}) to reconstruct the total wavelet coefficients $\omega_{E,B}$. An estimate of the polarization intensity of the source can also be obtained by looking at the central point of the combined coefficients, since
\begin{equation}\label{eq:central pix combination wavelet coefficients}
\omega_{E,B}(\vec{0},\hat{\phi},R)=\frac{P}{8\pi^2}\frac{R^2}{\sigma^2+R^2}\cos2\Big(\phi-\hat{\phi}^{E,B}(R)\Big).
\end{equation}
Consequently, if the $\hat{\phi}^{E,B}$ estimate is unbiased such that $\phi-\hat{\phi}^{E,B}\approx 0$, then the polarization intensity can be simply recovered from
\begin{equation}\label{eq:p est}
\hat{P}^{E,B}(\hat{\phi},R)=8\pi^2\frac{\sigma^2+R^2}{R^2}\omega_{E,B}(\vec{0},\hat{\phi},R).
\end{equation} 
We count now with two independent estimates of the source's polarization angle and intensity, that, in principle, should give similar values for the actual polarization angle and intensity. However, this will not be the case when we apply the filter to the real microwave sky since the backgrounds present in E-modes, both galactic foregrounds \citep{PolarizedDustForegrounds} and the CMB itself, are known to be higher than those in B-modes. Therefore, the results coming from the filtering of B-mode maps are expected to lead to a more accurate estimate. The same would happen if we were to apply an equivalent methodology for parameter estimation in $Q$ and $U$ maps since the ratio of background intensities between $Q$ and $U$ maps and B-modes is similar to that between E-modes and B-modes. Another advantage of working in E- and B-mode maps is that, thanks to the directionality of the point-source profile, independent estimates of both, polarization angle, and intensity, can be obtained from the study of E-modes and/or B-modes alone. In contrast, a joint analysis of $Q$ and $U$ is necessary to obtain a single estimate of $P$ and $\phi$ since point-sources are radially symmetric in $Q$ and $U$ maps.\\

We could also provide a joint estimate of the polarization angle and intensity of the source by combining the information from E- and B-mode maps. Exploiting the $\omega_x^E(r,\xi\pm\pi/4,\phi)=\omega_y^B(r,\xi,\phi)$ and $\omega_y^E(r,\xi\pm\pi/4,\phi)=-\omega_x^B(r,\xi,\phi)$ $\pi/4$ rotation symmetries between wavelet coefficients, we could rotate the $x$ and $y$ E-mode coefficients to match those from B-modes, and then stack them to create two new effective joint coefficients $\omega_x^J$ and $\omega_y^J$. Going a step further, we could account for the aforementioned differences on background amplitude between E- and B-modes by weighting the sum of wavelet coefficients: 
\begin{align}
    \omega_y^J(\vec{r},R)  = & \frac{\alpha^E_x}{\alpha^E_x+\alpha^B_y}  \omega^E_x(r,\xi+\pi/4, R) + \frac{\alpha^B_y}{\alpha^E_x+\alpha^B_y}   \omega^B_y(\vec{r},R), \notag\\
    \omega_x^J(\vec{r},R)  = &  - \frac{\alpha^E_y}{\alpha^B_x+\alpha^E_y}  \omega^E_y(r,\xi+\pi/4,R) + \frac{\alpha^B_x}{\alpha^B_x+\alpha^E_y} \omega^B_x(\vec{r},R), 
\end{align}
with indices reading $M=E,B$ and $i=x,y$. The $\alpha^M_i$ weights are defined from the dispersion of wavelet coefficients like $\alpha^M_i= \left(\sigma_{\omega^M_i}\right)^{-2}$. Using these weights we are giving more importance to the patches where the source's signal is best defined against the background. To properly calculate $\sigma_{\omega^M_i}$ and avoid the source to artificially boost the variance, we exclude an $8\sigma$ circular region around the source before computing the dispersion of $\omega^M_i$. \\

Since E-modes were rotated into B-modes to create $\omega^J$, these joint wavelet coefficients effectively behave like B-modes, and thus the joint estimates of the polarization angle and intensity should be computed like:
\begin{align}\label{eq:phi and p est joint}
\hat{\phi}^J(R)  = & \frac{1}{2} \arctan\left(\frac{-\omega_x^J(\vec{0,}R)}{\omega_y^J(\vec{0},R)}\right), \notag\\
\hat{P}^J(\hat{\phi},R) & =  8\pi^2\frac{\sigma^2+R^2}{R^2}\left(\cos 2\hat{\phi}^J\omega_y^J(\vec{0},R)-\sin 2\hat{\phi}^J\omega_x^J(\vec{0},R) \right).
\end{align}
By combining E- and B-modes at the wavelet coefficient stage, and then submitting them to the same parameter estimation logic, we ensure that polarization angles are correctly defined inside the $\phi\in[0,\pi)$ interval and that polarization intensities are always positive. Otherwise we could not guarantee the correct definition of our joint estimates.\\

\subsection{Calibration of pixelization effects}
\label{sec:calibration}
    
All the equations presented in previous sections rely on continuous functions. However, digital imaging discretizes information into pixels, compromising the resolution of functions to the number of pixels used. Therefore, a proper filter implementation must consider, and if necessary correct, the possible pixelization induced distortions.\\

\begin{table}[tbp]
\centering
\begin{tabular}{c|ccc|cc}
    \hline
    \multirow{2}{*}{$\nu$ /GHz} & \multicolumn{3}{c|}{PICO channels} & \multicolumn{2}{c}{Projected patches}   \\
    & nside & $L_{pix}$ /arcmin & FWHM /arcmin & $L_{patch}$ /deg & FWHM /pix\\
    \hline\hline
    30 & 512 & 6.87 & 28.3 & 7.33 & 4.119\\
    155 & 2048 & 1.72 & 6.2 & 1.83 & 3.609\\

\end{tabular}
\caption{\label{tab: pixelization parameters} Parameters defining the sphere pixelization and instrumental beam of the two PICO  frequency channels to simulate. We obtain the $L_{patch}$ side of the patches to project and the corresponding FWHM/pix ratio by fixing the patch extent to $64\times64$ pixels, and forcing pixels in the plane to have the same size as pixels in the sphere.}
\end{table}

Since our filter is implemented in the plane, the first step of the filtering process would be to project the target region of the sky (i.e., part of a spherical surface) onto the Cartesian plane. To guarantee that no significant distortions are introduced to the shape of sources by projecting, projections must be limited to a close entourage of the source, and pixels in the plane must have a similar (or smaller) size than pixels on the sphere. Given how the finesse of the sphere pixelization depends on the particular resolution granted by the instrumental specifications of every experiment, the plane's pixelization has to be tailored for the analysis of the specific data at hand. In particular, in this work we will be testing the performance of the filter on PICO-like simulations, so our pixelization is designed to fit their current instrumental specifications given in \cite{PICO}. Table \ref{tab: pixelization parameters} collects the main parameters describing both the sphere and plane pixelization of the two PICO channels we will be simulating. \\

Like most CMB experiments, PICO maps will be build using \verb|HEALPix|\footnote{\url{http://healpix.sourceforge.net}}, the \textit{Hierarchical Equal Area, and Iso-Latitude Pixelation} of the sphere proposed by \cite{healpix}. In this pixelization scheme, the sphere is divided into $12\times \mathrm{nside}^2$ rhomboid pixels. By fixing the patch extent to $64\times64$ pixels to keep the computational cost of the filtering process at bay, and forcing pixels in the plane to have the same size as pixels in the sphere, the size of the square regions to project is immediately set to be of $7.33^\circ\times7.33^\circ$ and $1.83^\circ\times1.83^\circ$, respectively, for the 30 GHz and 155 GHz channels. Patches of that size are large enough to offer a good representation of the statistical properties of background emissions, and small enough to ensure that the flat approximation of the sphere's surface will still hold, thus avoiding the introduction of distortions during projection.\\ 

\begin{figure}[tbp]
    \begin{center}
        \includegraphics[width=0.45\textwidth]{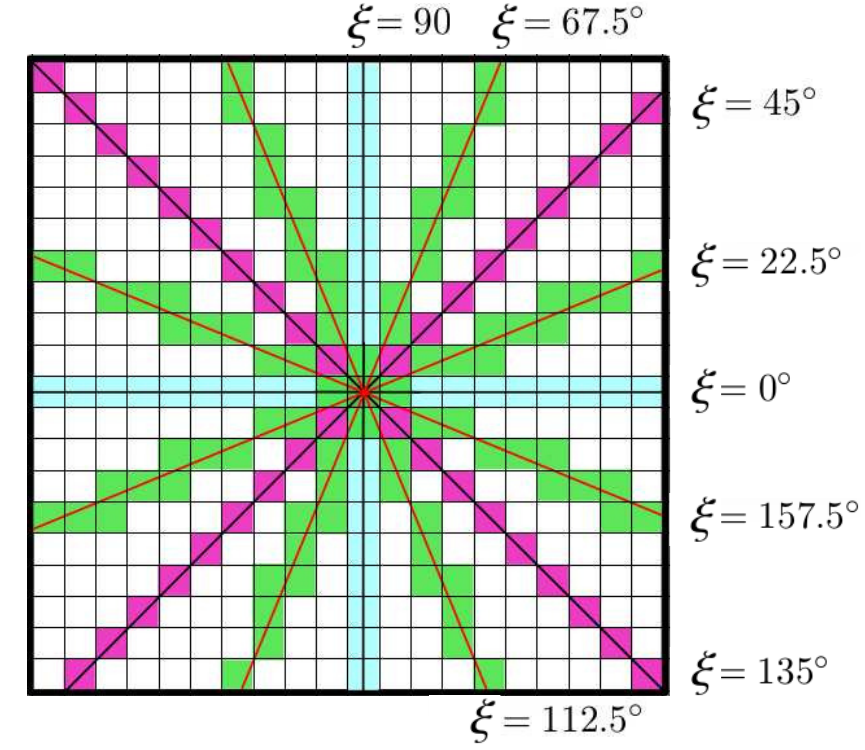}        
        \caption{\footnotesize The angular resolution of functions is limited by the pixel grid. Only sources with their lobes falling in the $x$ or $y$ axes or the diagonals will suffer no distortion. Due to the pixel grid, all other orientations will not be well defined.}
        \label{PixGrid}
    \end{center}
\end{figure} 

\begin{figure}[tbp]
    \begin{center}
        \includegraphics[width=1.0\textwidth]{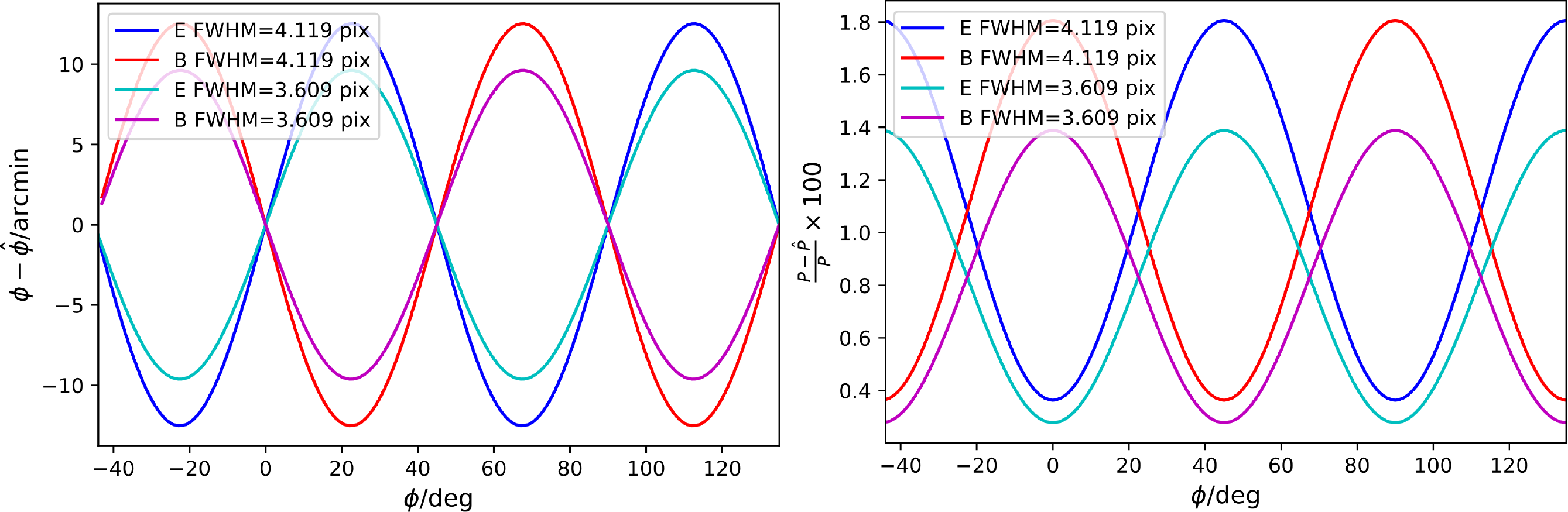}   
        \caption{\footnotesize Biases introduced by the chosen pixelization and FWHM/pix ratio in the determination of both polarization angle (left) and intensity (right), when applying the filter to a naked source. }
        \label{ErrNkdSrc}
    \end{center}
\end{figure}

As shown in figure \ref{PixGrid}, pixelization limits the angular resolution of the source's profile. The compact nature of the source limits its extension to the smallest of $r$s, where no matter how fine pixelization is, only the $\xi=\{0^\circ,45^\circ,90^\circ,135^\circ,180^\circ,225^\circ,270^\circ,315^\circ\}$ angles will be perfectly defined. In contrast, the angles right between those would be the ones most affected by pixelization distortions. Therefore, it is only for the $\phi=\{0^\circ,45^\circ,90^\circ,135^\circ\}$ polarization angles (when the lobes of point-sources fall along the direction of the $x$ and $y$ axes and the diagonals) that the filter will be free of bias in the estimation of $\hat{\phi}$, while for the angles just in the middle, $\phi=\{22.5^\circ,67.5^\circ,112.5^\circ,157.5^\circ\}$, the largest biases are expected. These pixelization imposed restrictions clearly manifest themselves in the determination of the polarization angle when applying the filter to a naked source, as shown in figure \ref{ErrNkdSrc}.\\

Since the estimation of polarization intensity depends on the accuracy of the $\hat{\phi}$ estimate like $\hat{P}\propto \cos 2(\phi-\hat{\phi})$, in turn, the largest biases in the recovered polarization intensity will be shifted to $\phi=\{0^\circ,45^\circ,90^\circ,135^\circ\}$. In addition, the accuracy in the determination of $P$ is also limited by how well the discrete points in the pixel grid can sample continuous functions. Naively modeling this discrepancy as $\hat{P}=P-\epsilon$, where the value of $\epsilon$ progressively decreases for finer pixel grids, and the bias in the polarization angle estimate simply as $A\cos 2\phi$, like the left panel of figure \ref{ErrNkdSrc} suggests we can do, the relative bias committed in the determination of the polarization intensity would behave as:
\begin{equation}
\frac{P-\hat{P}}{P}\propto 1-\left(1-\frac{\epsilon}{P}\right)\cos (2A\cos 2\phi).
\end{equation}
Since $\epsilon /P$ is very small but different from zero, this toy model explains why the relative bias in the determination of the polarization intensity seen in figure \ref{ErrNkdSrc} shockingly does not oscillate around zero. Moreover, giving $A$ and $\epsilon$ the actual values they present in these scenarios, the model precisely reproduces the relative biases displayed for the discrete polarization intensity.\\

A finer pixelization allows for both, a better angular resolution, and a more precise approximation of the value of continuous functions at all points, decreasing the induced biases in the polarization angle and intensity. Once pixelization is fixed, the only free parameter altering the resolution of the source's profile is the FWHM/pix ratio. An increase in the FWHM/pix ratio has the effect of smoothing the profile of the source. As variations in the value of the source's profile are now smaller from pixel to pixel, the ability of the filter to distinguish from one polarization angle to another is also diminished. Therefore, increasing the FWHM/pix ratio aggravates the biases committed in polarization angle and intensity determination, as can be seen in figure \ref{ErrNkdSrc}. Albeit not illustrated here, playing with the $R$ filter scale has the same effect that increasing or decreasing the FWHM/pix ratio.\\

Since these biases in polarization angle and intensity determination are exclusively caused by known parameters of image pixelization, filter definition and instrument resolution, they can be easily corrected. In our case a multiplicative calibration function would suffice to correct the initial estimations of $\hat{P}$ and $\hat{\phi}$. We can obtain such calibration functions from the initial outputs recovered when applying the filter to a naked source like:
\begin{align}\label{eq:funciones calibracion}
f_{E,B}(\hat{\phi},\mathrm{FWHM/pix},R/\sigma)=&\frac{\phi}{\hat{\phi}_{E,B}(\phi,\mathrm{FWHM/pix},R/\sigma)}, \notag\\
g_{E,B}(\hat{\phi},\mathrm{FWHM/pix},R/\sigma)=&\frac{P}{\hat{P}_{E,B}(\phi,\mathrm{FWHM/pix},R/\sigma)}.
\end{align}

Calibration functions have been computed in this way for polarization angles $\phi\in[0^\circ,180^\circ)$, with a one degree step between them, and for the FWHM/pix and $R/\sigma$ ratios that will be used later to test the filter performance. For polarization angles not tabulated, the value of the calibration function is interpolated using cubic splines. Once stored, calibration functions are used to correct the filter initial response simply like $\tilde{\phi}_{E,B}= f_{E,B}\times\hat{\phi}_{E,B}$ and $\tilde{P}_{E,B}=g_{E,B}\times\hat{P}_{E,B}$. Since $\hat{\phi}^J$ and $\hat{P}^J$ joint estimates effectively behave like B-modes, the same calibration is applied to them. After calibration, the remaining residual errors are only due to numerical precision (of the order of $10^{-10}$arcsec and $10^{-14}\%$, respectively for $\tilde{\phi}$ and $\tilde{P}$). \\ 

\section{Test on simulations}
\label{sec:test_on_simulations}

With the filter defined and calibrated, we proceed now to test its performance on realistic simulations of the microwave sky, where sources are immersed in a background of CMB and galactic foreground emissions, and can also be hidden below instrumental noise. Before statistically characterizing the filter's performance in subsection \ref{sec:results_R=sigma}, we will first describe our simulations of the microwave sky in subsection \ref{sec:simulation_description}.\\

\subsection{Simulations description}
\label{sec:simulation_description}

\begin{figure}\CenterFloatBoxes
\begin{floatrow}
\ffigbox[\FBwidth]
{\caption{Histogram showing the dispersion of the B-mode foreground emission of the 768 $7.33^\circ\times7.33^\circ$ patches projected from the 30 GHz microwave sky, and their breakdown into the different regions.}\label{HistPatchDisp}}
{\includegraphics[width=0.45\textwidth]{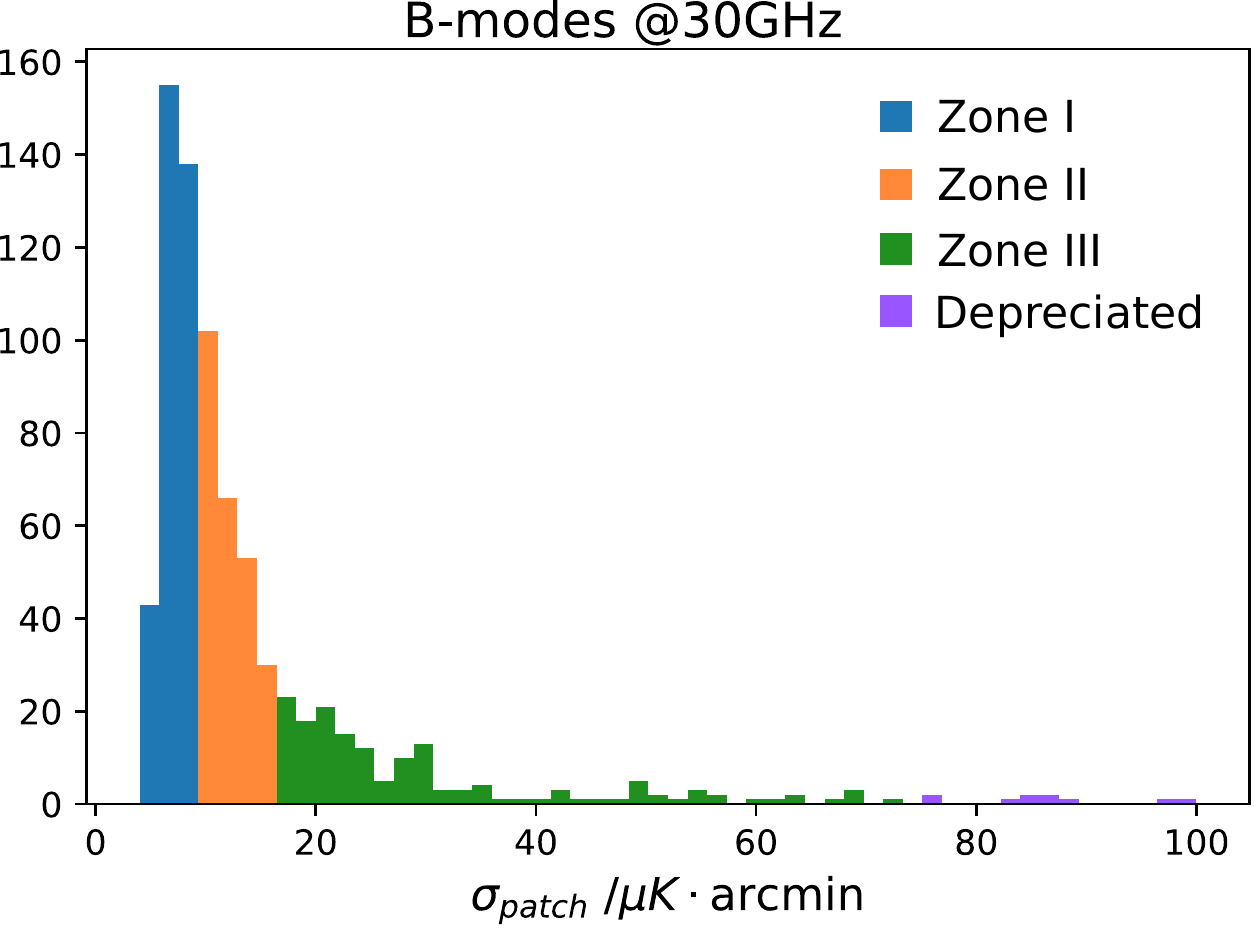}}
\killfloatstyle\capbtabbox[\Xhsize]
{\caption{Range of the projected patches' dispersion that defines the three representative regions of foreground emission, and CMB-only scenario, where the filter will be tested.  Note that the values of the CMB dispersion change from band to band because of the different pixel and FWHM resolution of each of them.}\label{RegionDef}}
{        \begin{tabular}{c c c c}
            \hline 
            & &\multicolumn{2}{c}{$\sigma_{\mathrm{patch}}$ /$\mu K\cdot$arcmin}\\
                & & 30GHz & 155 GHz\\
                \hline\hline
                E & CMB-only & [29.27, 34.02] & [8.76, 12.76] \\
                & Zone I & [30.11, 34.11) & [11.21, 23.68) \\   
                & Zone II & [34.11, 41.82) & [23.68, 72.75) \\
                & Zone III &  [41.82, 116.22) & [72.75, 596.70)\\
                \hline\hline
                B & CMB-only & [1.66, 1.87] & [0.52, 0.71] \\
                & Zone I & [4.46, 8.80) & [5.82, 20.90)  \\   
                & Zone II & [8.80, 15.79) & [20.90, 71.70)  \\
                & Zone III & [15.79, 72.51) & [71.70, 593.17) 
        \end{tabular}}
\end{floatrow}
\end{figure}

We decided to test the filter performance on simulations of the future PICO satellite, an ideal experiment for point-source detection since it will combine high resolutions with low instrumental noises. Amongst the 21 frequency bands envisioned in the \cite{PICO} mission concept study, we chose to work with the 30 GHz and 155 GHz channels. We selected these bands for the diverse experimental conditions they will allow us to explore: from different beam sizes ($28$ vs. $6$ arcmin, as indicated in table \ref{tab: pixelization parameters}), to contrasting backgrounds (see table \ref{RegionDef}). On the one hand, galactic foregrounds have a similar amplitude in E- and B-modes at 155 GHz, while at 30 GHz, the amplitude of galactic E-modes is larger than that of B-modes. On the other hand, B-modes will still be noise-dominated at 30 GHz, whereas, thanks to the low instrumental noise planned for the 155 GHz band, B-modes will be foreground-dominated at 155 GHz (see table \ref{Noise&Fluxes}).\\    

Foreground emission was simulated using the \textit{Planck Sky Model} \citep{PSM}, a publicly available software\footnote{\url{https://pla.esac.esa.int/\#plaavi_psm}} that allows us to generate random realizations of the microwave sky in agreement with current observational constraints. Our simulations include the lensed CMB, synchrotron and thermal dust emissions, and a background of faint point-sources below the detection threshold expected for PICO in intensity (4 mJy and 7 mJy respectively for the 30 GHz and 155 GHz channels \cite{PICO}). The CMB is simulated in concordance with the cosmological model obtained by the \emph{Planck} mission \cite{planck18_cosmologicalParams}, and assuming an $r=0$. For the simulation of Galactic foregrounds, both the spectral index of the power law defining synchrotron emission, and the dust temperature and spectral index of the modified blackbody emulating the thermal dust, vary across the sky according to experimental constraints \cite{planck2015foregrounds,PolarizedDustForegrounds}. The sources to characterize will be added directly on the plane once patches are projected. Realizations of isotropic white Gaussian noise of the typical noise levels expected for PICO (see table \ref{Noise&Fluxes}) will also be added directly on the plane once patches are projected.\\

We will simulate sources of different fluxes, ranging from tens of mJy to 10 Jy in intensity, as indicated in the $I_i$ column of table \ref{Noise&Fluxes}. The aim of this particular flux selection is not to make a full characterization of the filter performance for all fluxes above the detection threshold but rather to show the potential of the methodology through a few archetypal fluxes. To translate fluxes into polarization intensities, we will assume sources to have a constant $\Pi=0.02$ polarization degree, independent of their flux or the frequency band like \cite{Bonavera} and \cite{srcSPTPol} suggest. Accounting also for the conversion factor between intensity and thermodynamical units (more about unit conversion can be found in \cite{Hobson}), the polarization intensity of sources will be
\begin{equation}
     P(\mu K)\approx \Pi \hspace{1 mm} I(\mathrm{Jy}) \hspace{1 mm} \frac{\sinh^2 (x/2)}{24.8 x^4}\hspace{1mm}, \hspace{10 mm} x\approx \frac{\nu}{56.8 \mathrm{GHz}}.
\end{equation}

\begin{figure}[tbp]
    \begin{center}
        \makebox[\textwidth][c]{\includegraphics[width=1.2\textwidth]{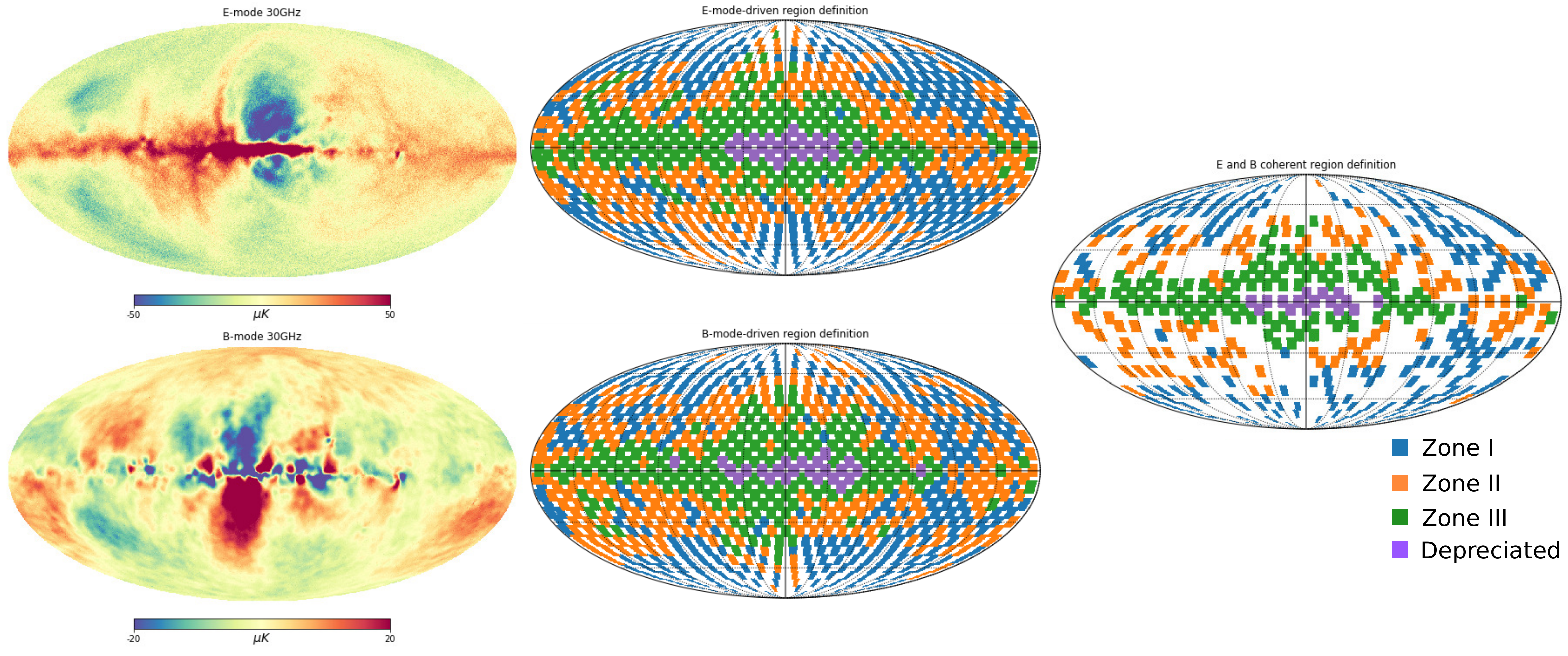}}
        \caption{\footnotesize Starting from the simulated 30 GHz sky in E- and B-modes (left column), we project 768 square patches of $7.33^\circ$ side centered around the positions of \texttt{HEALPix}'s $\mathrm{nside}=8$ pixels, and classify them into three different regions (center column) so that Zone I (blue) contains the first $40\%$ of patches of lowest dispersion, Zone II (orange) comprises the next $35\%$ of patches of lowest dispersion, and Zone III (green) collects the next $22\%$. The remaining $3\%$ of the patches (purple) are depreciated for corresponding to the regions of largest foreground emission inside the galactic plane. As an additional condition, we also require this classification to be spatially coherent across polarization modes (right column), meaning that for a certain patch to belong to a given region, both its dispersion in E- and B-modes needs to fall into that zone.}
        \label{regionsOnTheSky}
    \end{center}
\end{figure}

As can be seen in the left column of figure \ref{regionsOnTheSky} for the 30 GHz band, foreground emission greatly varies across the sky, so to better assess filter performance, we divided the sky into three separate regions based on the intensity of foreground emission. To ensure the sampling of all types of foreground emission across the sky, we start by projecting a total of 768 square patches (of a $7.33^\circ$ side for the 30 GHz channel, and $1.83^\circ$ for the 155 GHz one) centered around the positions of \verb|HEALPix|'s $\mathrm{nside}=8$ pixels. The three distinct regions are then defined as a function of the $\sigma_{\mathrm{patch}}$ dispersion foreground emission presents on each of those patches so that \textit{Zone I} contains the first $40\%$ of patches of lowest dispersion, \textit{Zone II} comprises the next $35\%$ of patches of lowest dispersion, and \textit{Zone III} collects the next $22\%$ (see figure \ref{HistPatchDisp} for an example). The remaining $3\%$ of the patches are depreciated for corresponding to the regions of largest foreground emission inside the galactic plane. This would lead us to an independent region definition for E- and B-modes for each frequency band (see the central column of figure \ref{regionsOnTheSky} for an example). As an additional condition, we require this classification to be spatially coherent across polarization modes, meaning that for a certain patch to belong to a given region, both its dispersion in E- and B-modes needs to fall into that zone. As can be seen in the right column of figure \ref{regionsOnTheSky}, enforcing this condition can severely reduce the number of patches contained in each zone. Nevertheless, after applying this constraint, we still count with more than a hundred patches per zone to use for the statistical characterization of the filter performance in section \ref{sec:results_R=sigma}.\\ 

We could also impose a spatial coherence across frequencies, but that would be of little use given the very different nature of foregrounds in the two frequency bands: the 30 GHz band is mainly composed of synchrotron radiation, while thermal dust emission dominates on the 155 GHz band. For this reason, regions are defined independently for each frequency. Finally, table \ref{RegionDef} shows the $\sigma_{\mathrm{patch}}$ range defining the three regions like so constructed, both for E- and B-modes, and the two frequency bands. The dispersion of the simulated CMB component is also included for reference.\\

\begin{table}[tbp]
\centering
\begin{tabular}{c|cccc}
    \hline
    $\nu$ /GHz & $I_i$ /Jy & $P_i$ /pK & $\sigma_n$ /$\mu K\cdot$arcmin \\
    \hline\hline
    30 & 0.07, 0.5, 1, 10 & 52, 370, 740, 7403 & 12.4\\
    155 & 0.04, 0.5, 1, 10 & 2, 24, 49, 487 & 1.8
\end{tabular}
\caption{ Different fluxes, both in intensity and polarization intensity, of the sources simulated for the two PICO frequency bands. In what is left of the paper, polarization intensities will be identified by the $i$ sub-index following the order of appearance in this table. The instrumental noise of each channel is also provided.}
\label{Noise&Fluxes}
\end{table}

\subsection{Statistical characterization of the filter performance}
\label{sec:results_R=sigma}

To characterize the filter performance, we randomly select a hundred patches from each of the regions defined in table \ref{RegionDef}, and add to them a realization of isotropic Gaussian noise. At the center of each patch, we place a source directly on the plane for 36 different polarization angles, and the four fluxes shown in table \ref{Noise&Fluxes}. By applying the filter to all of them, we obtain 3600 estimates of $\tilde{\phi}$ and $\tilde{P}$ for the statistical analysis of the filter performance for each flux and region.\\

\begin{figure}[tbp]
    \begin{center}
        \includegraphics[width=1.0\textwidth]{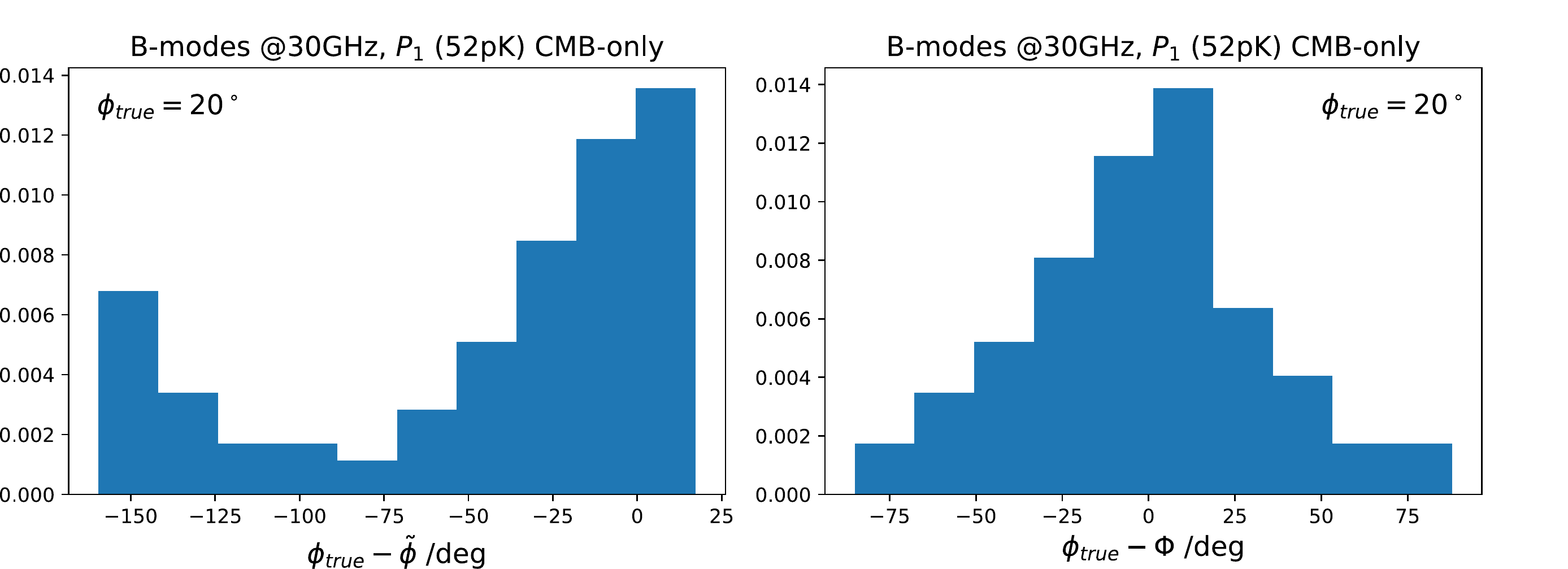}        
        \caption{\footnotesize The periodic and bounded nature of the polarization angle can break the continuity of the distribution of recovered $\tilde{\phi}$ values around the true input value (left) since part of the recovered angles will be $\pm\pi$ shifted. In this way, if we were to evaluate the uncertainty in polarization angle estimation directly from the distribution of $\tilde{\phi}$, the dispersion of recovered values for angles near the edges of the definition interval will be artificially increased because of this periodicity effect. We can work instead with the $\Phi=\{\tilde{\phi}, \tilde{\phi}-\pi, \tilde{\phi}+\pi\}$ angle that minimizes the $|\phi-\Phi|$ difference (rigth). Since $\Phi$ does take into account the periodicity of the polarization angle, its values present a Gaussian-like distribution around the true input angle, and therefore, we can quantify the $\sigma_\phi$ uncertainty as the standard deviation of $\Phi$ values.}
        \label{histograms_angle}
    \end{center}
\end{figure}

However, the intrinsic nature of the source's parameters complicates their statistical analysis. On the one hand, the polarization angle is a bounded ($\phi\in[0,\pi)$) and periodic (the source's profile is symmetric under a $\phi\pm\pi$ rotation) quantity. This means that for $\phi^*$ angles close to the edges of the definition interval, the filter is equally likely to return $\phi^*\pm\pi$ since both of them are actually equivalent and angles outside the definition interval are not allowed. This feature does not suppose a problem in the sense that we know errors should also be interpreted periodically (i.e., a $\phi^*=5^\circ\pm10^\circ$ uncertainty means that $\phi^*$ is compatible with all angles contained in the $[0^\circ,15^\circ]\cup[175^\circ,180^\circ)$ interval), but, as can be seen in the left panel of figure \ref{histograms_angle}, it will break the continuity of the distribution of recovered values around the true input value since part of the recovered angles will be $\pm\pi$ shifted. In this way, if we were to evaluate the uncertainty in polarization angle estimation directly from the distribution of recovered $\tilde{\phi}$, the periodicity of $\phi$ will artificially increase the dispersion of recovered values for angles near the edges of the definition interval.\\

To solve this problem, we work with the $\Phi=\{\tilde{\phi}, \tilde{\phi}-\pi, \tilde{\phi}+\pi\}$ angle that minimizes the $|\phi-\Phi|$ difference instead of with the $\tilde{\phi}$ estimate itself. Since this new $\Phi$ does take into account the periodicity of the polarization angle, its values present a Gaussian-like distribution around the true angle (see the right panel of figure \ref{histograms_angle}), and therefore, we can quantify the $\sigma_\phi$ uncertainty in polarization angle recovery as the standard deviation of the values recovered for all patches, averaged over the different orientations of the source.\\

\begin{figure}[tbp]
    \begin{center}
        \includegraphics[width=1.0\textwidth]{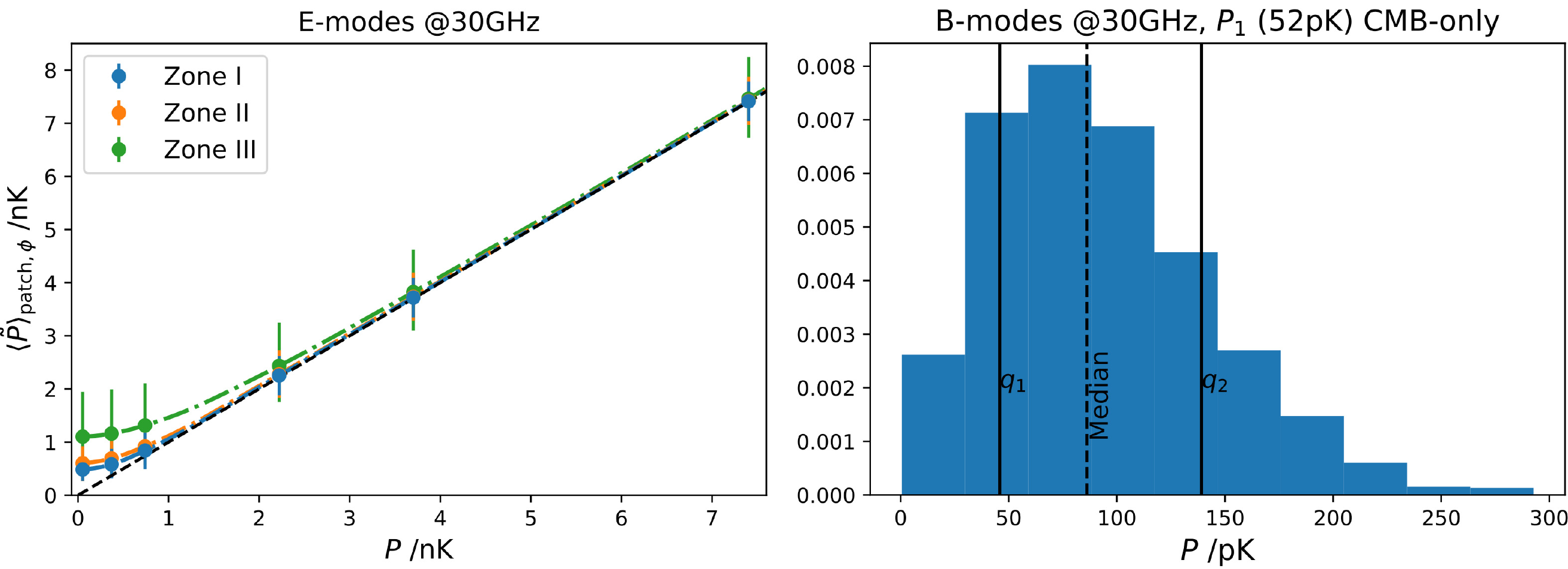}        
        \caption{\footnotesize \emph{Left)} Mean polarization intensity recovered (averaged over all patches, and over all orientations of the source) against the true polarization intensity of the sources. The distinctive plateau the recovered polarization intensity shows at low values is an indicative of the polarization intensity value below which sources can no longer be correctly characterized. \emph{Right)} Because it is a positive defined quantity, the distribution of recovered $\tilde{P}$ values is skewed towards larger polarization intensities for the faintest of sources. For this reason, we give asymmetric $\sigma_P^{\mathrm{lower}}$ and $\sigma_P^{\mathrm{upper}}$ error bars corresponding to the difference of the $q_1=0.175$ and $q_2=0.825$ quantiles with the median, and the $b_P$ bias as an estimation of how much the filter tends to overestimate polarization intensity.}
        \label{POverEst}
    \end{center}
\end{figure}

On the other hand, polarization intensity is a positive defined quantity (i.e., $P>0$ always). This also restricts the interval of allowed values, and for the faintest of sources, it has the effect of skewing the distribution of recovered $\tilde{P}$ towards larger polarization intensities (see the right panel of figure \ref{POverEst}). Therefore, there would be a limiting value of $P$ below which the filter will start to systematically overestimate polarization intensity, as can be seen in the left panel of figure \ref{POverEst}, where the mean $\tilde{P}$ recovered is plotted against the actual polarization intensity. For this reason, to properly characterize its performance, we should also give a measurement of how much the filter tends to overestimate polarization intensity. The parameter we chose to quantify this is the $b_P$ bias, calculated as the mean error in the estimation of $P$, averaged over all patches, and over all orientations of the source: $b_P=\langle \tilde{P}-P \rangle_{\mathrm{patch,}\phi}$. Since the distribution of $\tilde{P}$ estimates is asymmetric, we give asymmetric $\sigma_P^{\mathrm{lower}}$ and $\sigma_P^{\mathrm{upper}}$ error bars corresponding to the difference of the $q_1=0.175$ and $q_2=0.825$ quantiles with the median (see the right panel of figure \ref{POverEst}). As a general rule, sources will only be correctly characterized as long as $b_P<\sigma_P^{\mathrm{lower}}$.\\

Taking into account all these peculiarities, tables \ref{BaseResutls30GHz} and \ref{BaseResutls155GHz} collect the typical errors (presented in a $\sigma_\phi$; $b_{P\hspace{5pt}-\sigma_{P}^{\mathrm{lower}}}^{+\sigma_P^{\mathrm{upper}}}$ format) obtained for each frequency band, sky region and polarization intensity tested. Let us focus on the results for the 30 GHz band. For the lowest of polarization intensities tested ($P_1$ is only ten times above the detection threshold in intensity), the filter is not able to characterize point-sources in any of the sky regions. Increasing the polarization intensity up to $P_2$ (corresponding to a $500$ mJy intensity), characterization starts to be possible for B-modes in the Zones I and II of low foreground emission. Although still high, biases also improve for E-modes, reaching $b_P<\sigma_P^{\mathrm{lower}}$ values. For sources of $P_3$ polarization intensity, characterization is now possible in all sky regions for B-modes, and only in the Zones I and II of low foreground amplitude for E-modes. As would be expected, characterization is possible in all regions, and in both E- and B-modes, for very bright sources ($P_4$ corresponds to a $10$ Jy intensity).\\

\begin{table}[tbp]
\begin{center}
    \begin{tabular}{c c c c c c}
    \hline  
    \multicolumn{2}{c}{\multirow{2}{*}{$30$ GHz}} & $P_{1}$ & $P_{2}$ & $P_{3}$ & $P_{4}$ \\
    & & (52 pK/ 0.07 Jy) & (370 pK/ 0.5 Jy) & (740 pK/ 1 Jy) & (7403 pK/ 10 Jy)\\
        \hline\hline
        \multirow{2}{*}{CMB-only} & E &  48.5 ; 362 $^{+226}_{\hspace{7pt}-206}$ & 30.0 ; 163 $^{+281}_{\hspace{7pt}-222}$ & 14.9 ;  77 $^{+308}_{\hspace{7pt}-284}$ & 1.3 ;   7 $^{+313}_{\hspace{7pt}-312}$\\
        & B & 34.9 ;  41 $^{+ 53}_{\hspace{7pt}- 40}$ & 5.1 ;   6 $^{+ 59}_{\hspace{7pt}- 56}$ & 2.5 ;   3 $^{+ 58}_{\hspace{7pt}- 58}$ & 0.3 ;   0 $^{+ 58}_{\hspace{7pt}- 59}$ \\
        & J & 34.4 ;  41 $^{+ 57}_{\hspace{7pt}- 38}$ & 5.1 ;   6 $^{+ 57}_{\hspace{7pt}- 55}$ & 2.5 ;   3 $^{+ 57}_{\hspace{7pt}- 56}$ & 0.4 ;   2 $^{+101}_{\hspace{7pt}-105}$ \\
        \hline
        \multirow{2}{*}{Zone I} & E &  49.5 ; 433 $^{+214}_{\hspace{7pt}-217}$ & 33.4 ; 212 $^{+293}_{\hspace{7pt}-265}$ & 17.3 ; 101 $^{+336}_{\hspace{7pt}-351}$ & 1.4 ;   9 $^{+369}_{\hspace{7pt}-373}$ \\
        & B & 44.7 ; 147 $^{+101}_{\hspace{7pt}- 96}$ & 13.8 ;  34 $^{+144}_{\hspace{7pt}-135}$ & 6.1 ;  16 $^{+149}_{\hspace{7pt}-143}$ & 0.6 ;   2 $^{+148}_{\hspace{7pt}-148}$\\
        & J &  44.6 ; 150 $^{+104}_{\hspace{7pt}-106}$ & 14.1 ;  34 $^{+153}_{\hspace{7pt}-133}$ & 6.3 ;  14 $^{+158}_{\hspace{7pt}-144}$ & 0.6 ; -14 $^{+173}_{\hspace{7pt}-160}$\\
        \hline
        \multirow{2}{*}{Zone II} & E & 49.6 ; 551 $^{+319}_{\hspace{7pt}-338}$ & 36.3 ; 323 $^{+392}_{\hspace{7pt}-295}$ & 23.1 ; 180 $^{+446}_{\hspace{7pt}-349}$ & 1.9 ;  16 $^{+456}_{\hspace{7pt}-449}$\\
        & B &  47.2 ; 210 $^{+140}_{\hspace{7pt}-111}$ & 19.5 ;  63 $^{+182}_{\hspace{7pt}-175}$ & 8.3 ;  29 $^{+188}_{\hspace{7pt}-187}$ & 0.8 ;   3 $^{+191}_{\hspace{7pt}-191}$\\
        & J & 46.9 ; 221 $^{+160}_{\hspace{7pt}-121}$ & 20.8 ;  70 $^{+191}_{\hspace{7pt}-171}$ & 9.2 ;  29 $^{+195}_{\hspace{7pt}-190}$ & 0.8 ; -31 $^{+196}_{\hspace{7pt}-196}$\\
        \hline
        \multirow{2}{*}{Zone III} & E & 50.6 ; 1051 $^{+843}_{\hspace{7pt}-480}$ & 42.2 ; 790 $^{+833}_{\hspace{7pt}-520}$ & 33.7 ; 574 $^{+793}_{\hspace{7pt}-541}$ & 3.7 ;  59 $^{+786}_{\hspace{7pt}-737}$\\
        & B & 49.6 ; 569 $^{+420}_{\hspace{7pt}-196}$ & 34.3 ; 345 $^{+397}_{\hspace{7pt}-300}$ & 22.2 ; 219 $^{+389}_{\hspace{7pt}-349}$ & 2.3 ;  23 $^{+391}_{\hspace{7pt}-388}$\\
        & J & 49.2 ; 530 $^{+457}_{\hspace{7pt}-242}$ & 33.9 ; 312 $^{+417}_{\hspace{7pt}-274}$ & 22.4 ; 186 $^{+410}_{\hspace{7pt}-316}$ & 2.0 ; -38 $^{+377}_{\hspace{7pt}-386}$\\
    \end{tabular}
\end{center}
\caption{\footnotesize  Typical $\sigma_\phi$; $b_{P\hspace{5pt}-\sigma_{P}^{\mathrm{lower}}}^{+\sigma_P^{\mathrm{upper}}}$ errors in the determination of the polarization angle (in degrees) and intensity (in pK) for the different scenarios considered at the 30 GHz PICO-like band. By its definition, a $b_P>0$ indicates an overestimation of $P$, while $b_P<0$ corresponds to an underestimate. A zero bias means that the value of $b_P$ falls below the chosen precision for this table. As reference, the value of true input polarization intensity is shown in the header of each column.}
\label{BaseResutls30GHz}
\end{table}

\begin{table}[tbp]
\begin{center}
    \begin{tabular}{c c c c c c}
    \hline 
        \multicolumn{2}{c}{\multirow{2}{*}{$155$ GHz}} & $P_{1}$  & $P_{2}$  & $P_{3}$  & $P_{4}$ \\
        &  & (2 pK/ 0.04 Jy) & (24 pK/ 0.5 Jy) &  (49 pK/ 1 Jy) & (487 pK/ 10Jy)\\
        \hline\hline
        \multirow{2}{*}{CMB-only} & E & 51.1 ;  68 $^{+ 35}_{\hspace{7pt}- 28}$ & 43.6 ;  48 $^{+ 39}_{\hspace{7pt}- 31}$ & 34.4 ;  33 $^{+ 43}_{\hspace{7pt}- 37}$ & 3.2 ;   3 $^{+ 51}_{\hspace{7pt}- 51}$\\
        & B & 40.8 ;   3 $^{+  2}_{\hspace{7pt}-  2}$ & 4.1 ;   0 $^{+  3}_{\hspace{7pt}-  3}$ & 2.0 ;   0 $^{+  3}_{\hspace{7pt}-  3}$ & 0.2 ;   0 $^{+  3}_{\hspace{7pt}-  3}$\\
        & J & 41.0 ;   3 $^{+  2}_{\hspace{7pt}-  2}$ & 4.1 ;   0 $^{+  3}_{\hspace{7pt}-  3}$ & 2.0 ;   0 $^{+  3}_{\hspace{7pt}-  3}$ & 0.3 ;  -0 $^{+  5}_{\hspace{7pt}-  5}$\\
        \hline
        \multirow{2}{*}{Zone I} & E & 51.2 ;  66 $^{+ 42}_{\hspace{7pt}- 27}$ & 43.0 ;  47 $^{+ 37}_{\hspace{7pt}- 33}$ & 34.1 ;  32 $^{+ 41}_{\hspace{7pt}- 36}$ & 3.2 ;   3 $^{+ 51}_{\hspace{7pt}- 50}$\\
        & B & 49.0 ;  20 $^{+ 17}_{\hspace{7pt}-  8}$ & 24.5 ;   8 $^{+ 16}_{\hspace{7pt}- 11}$ & 13.5 ;   4 $^{+ 16}_{\hspace{7pt}- 13}$ & 1.1 ;   0 $^{+ 14}_{\hspace{7pt}- 14}$\\
        & J &  49.3 ;  18 $^{+ 13}_{\hspace{7pt}-  7}$ & 22.8 ;   6 $^{+ 14}_{\hspace{7pt}- 12}$ & 11.3 ;   3 $^{+ 14}_{\hspace{7pt}- 13}$ & 1.0 ;  -0 $^{+ 14}_{\hspace{7pt}- 14}$\\
        \hline
        \multirow{2}{*}{Zone II} & E & 51.5 ;  96 $^{+ 43}_{\hspace{7pt}- 44}$ & 45.7 ;  76 $^{+ 46}_{\hspace{7pt}- 44}$ & 39.6 ;  58 $^{+ 54}_{\hspace{7pt}- 47}$ & 4.6 ;   6 $^{+ 74}_{\hspace{7pt}- 76}$\\
        & B & 50.9 ;  62 $^{+ 53}_{\hspace{7pt}- 24}$ & 41.2 ;  44 $^{+ 50}_{\hspace{7pt}- 31}$ & 31.6 ;  31 $^{+ 48}_{\hspace{7pt}- 32}$ & 3.2 ;   3 $^{+ 45}_{\hspace{7pt}- 43}$\\
        & J &  51.1 ;  50 $^{+ 45}_{\hspace{7pt}- 18}$ & 38.9 ;  32 $^{+ 40}_{\hspace{7pt}- 24}$ & 28.1 ;  21 $^{+ 36}_{\hspace{7pt}- 30}$ & 2.5 ;   1 $^{+ 35}_{\hspace{7pt}- 35}$\\
        \hline
        \multirow{2}{*}{Zone III} & E & 51.8 ; 373 $^{+269}_{\hspace{7pt}-191}$ & 49.6 ; 352 $^{+265}_{\hspace{7pt}-187}$ & 47.3 ; 330 $^{+269}_{\hspace{7pt}-186}$ & 21.5 ; 131 $^{+286}_{\hspace{7pt}-178}$\\
        & B &  51.8 ; 347 $^{+336}_{\hspace{7pt}-124}$ & 49.5 ; 325 $^{+339}_{\hspace{7pt}-122}$ & 46.8 ; 304 $^{+334}_{\hspace{7pt}-128}$ & 20.6 ; 124 $^{+246}_{\hspace{7pt}-162}$\\
        & J & 51.7 ; 246 $^{+139}_{\hspace{7pt}-120}$ & 48.6 ; 225 $^{+141}_{\hspace{7pt}-121}$ & 45.4 ; 205 $^{+151}_{\hspace{7pt}-117}$ & 15.6 ;  58 $^{+179}_{\hspace{7pt}-138}$\\
    \end{tabular}
\end{center}
\caption{\footnotesize  Typical $\sigma_\phi$; $b_{P\hspace{5pt}-\sigma_{P}^{\mathrm{lower}}}^{+\sigma_P^{\mathrm{upper}}}$ errors in the determination of the polarization angle (in degrees) and intensity (in pK) for the different scenarios considered at the 155 GHz PICO-like band. By its definition, a $b_P>0$ indicates an overestimation of $P$, while $b_P<0$ corresponds to an underestimate. A zero bias means that the value of $b_P$ falls below the chosen precision for this table. As reference, the value of true input polarization intensity is shown in the header of each column. } 
\label{BaseResutls155GHz}
\end{table}

\begin{table}[tbp]
\begin{center}
    \begin{tabular}{c c c c | c c c | c c c | c c c}
    \hline 
    \multirow{2}{*}{$\nu$ /GHz}&\multicolumn{3}{c|}{CMB-only} & \multicolumn{3}{c|}{Zone I} & \multicolumn{3}{c|}{Zone II} & \multicolumn{3}{c}{Zone III}\\
         & E & B & J & E & B & J & E & B & J & E & B & J \\
        \hline\hline
         30 & 270 & 55 & 62 & 297 & 132 & 125 & 419 & 169 & 180 & 800 & 435 & 436\\
         155 & 44 & 5 & 5 & 42 & 18 & 14 & 64 & 45 & 34 & 356 & 335 & 215
    \end{tabular}
\end{center}
\caption{\footnotesize Polarization intensity of extragalactic sources (in pK) below which $b_P>\sigma_P^{\mathrm{lower}}$, and thus sources cannot be correctly characterized.} 
\label{DetThres}
\end{table}

A similar pattern can be seen on the 155 GHz band, although this time, a good characterization of point-sources emission is harder to reach because of the larger foreground emission present on this band (see table \ref{RegionDef}). In this case, apart from very bright sources, the filter is only able to correctly characterize sources of $P_2$ and $P_3$ polarization intensities in the region of lowest foreground emission, and only in the B-mode channel.\\

\begin{figure}[tbp]
    \begin{center}
        \includegraphics[width=1.0\textwidth]{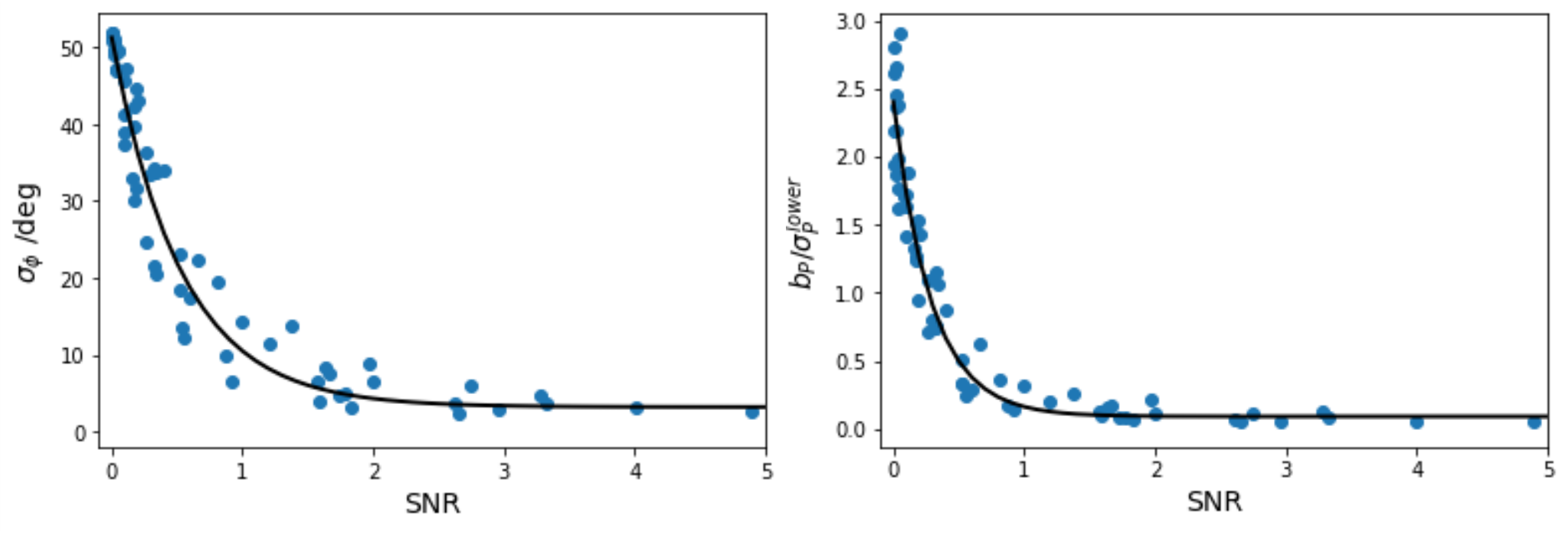}        
        \caption{\footnotesize  $\sigma_\phi$ (left) and $b_P/\sigma_P^{\mathrm{lower}}$ (right) errors from tables \ref{BaseResutls30GHz} and \ref{BaseResutls155GHz} as a function of the signal-to-noise ratio of the source with respect to the background, where signal-to-noise is computed as the maximum amplitude of the point-source in E- and B-mode maps (the value the source takes at the peak of its positive lobes) over the mean $\sigma_{\mathrm{patch}}$ dispersion of the patches of that region of the sky. With higher signal-to-noise ratios, errors decrease following, approximately, a $a+be^{-x/c}$ law (black lines).}
        \label{errVSsnr}
    \end{center}
\end{figure}

In both bands, all four $\sigma_\phi$, $b_P$, $\sigma_P^{\mathrm{lower}}$ and $\sigma_P^{\mathrm{upper}}$ parameters prove to be smaller for B-modes than E-modes, with the differences between both modes decreasing as the amplitude of foreground emission increases. It is only for Zone III of the 155 GHz band, where foregrounds become equally important for both modes (see table \ref{RegionDef}), that working on B-modes does not report an obvious advantage compared to E-modes. Such results confirm our initial hypothesis that B-mode polarization maps would be the best channel for the study of point-sources because of the lower amplitude of the backgrounds found there.\\

Meanwhile, the joint parameter estimation tends to return very similar results to those coming from B-modes at the 30 GHz band since the lower amplitude of foregrounds there makes B-modes the main contributor to $\hat{\phi}$ and $\hat{P}$. In contrast, for the 155 GHz band, where the amplitude of foreground emission in E- and B-modes becomes comparable, the joint analysis starts to systematically yield better results than what a B-mode-only analysis would. The small negative biases recovered in some regions for the highest of fluxes reflect how in those scenarios $P$ is so far from zero that the recovered $\hat{P}$ values can be symmetrically distributed around the true $P$, and thus underestimating the polarization intensity becomes a possibility again.\\

For regions of low signal-to-noise ratio, where sources are too faint to stand above galactic foreground emission (e.g., sources of $P_2$ intensity on Zone III, or sources of $P_1$ on all regions of the sky), the filter is unable to characterize the emission of point-sources. In those scenarios, background emission dominates the filtering process, leading to a biased estimation of the polarization intensity and a poorly constrained estimate of the polarization angle. In this way, the information in tables \ref{BaseResutls30GHz} and \ref{BaseResutls155GHz} can be rearranged to be interpreted in terms of the signal-to-noise of the sources with respect to the background, so that these results can be extrapolated to any other region of the sky, frequency, or polarization intensity. This is what we do in figure \ref{errVSsnr}, where the $\sigma_\phi$ and $b_P/\sigma_P^{\mathrm{lower}}$ errors in each cell of tables \ref{BaseResutls30GHz} and \ref{BaseResutls155GHz} are presented as a function of the typical signal-to-noise ratio for that polarization intensity and region of the sky. The signal-to-noise ratio is computed as the maximum amplitude of the point-source in E- and B-mode maps (the value the source takes at the peak of its positive lobes) over the mean $\sigma_{\mathrm{patch}}$ dispersion of the patches of that region. As can be seen in figure \ref{errVSsnr}, the filter is able to characterize the polarization intensity of the source ($b_P\leq\sigma_P^{\mathrm{lower}}$ condition) as long as the source presents an $SNR\geq0.25$, while, for a good characterization of the polarization angle ($\sigma_\phi<5^\circ$), an $SNR\geq1.7$ is needed. \\

As a final summary of this statistical study of the filter performance, we give in table \ref{DetThres} the polarization intensity below which the filter will not be able to correctly characterize the properties of extragalactic sources. This threshold is obtained by finding the polarization intensity at which $b_P=\sigma_P^{\mathrm{lower}}$ for each region of the sky, which is equivalent to finding the point in figure \ref{POverEst} where $\langle \hat{P}\rangle-\sigma_P^{\mathrm{lower}}$ intersects the $P=P_{\mathrm{true}}$ line, and/or finding the polarization intensity that grants enough $SNR$ with respect to the background to reach the $b_P/\sigma_P^{\mathrm{lower}}=1$ level in figure \ref{errVSsnr}. Again, these results evidence how a better characterization is possible in maps of the B-mode polarization.\\


\section{Conclusions and future work}
\label{sec:conclusions}

In this work we have designed a filter based on steerable wavelets that allows the characterization of extragalactic point-sources on the E- and B-mode maps of the CMB polarization. The initial motivation for working in E- and B-modes maps instead of following the conventional approach of working in maps of the Stokes' $Q$ and $U$ parameters, where sources have a simpler profile, was to try to exploit the lower amplitude that the background of microwave emissions present in B-modes. The application of the filter to realistic simulations of the microwave sky proved that, indeed, a better determination of the sources' properties was possible in B-modes than in E-modes. Moreover, since a similar ratio of background amplitudes can be found between E- and B-modes, and between $Q$ and $U$ maps and B-modes, our results also prove that B-modes are the optimum polarization channel for point-source characterization.\\

Throughout this work, the filter scale has always been fixed to match the size of the source ($R=\sigma$). Nevertheless, filter performance could be enhanced by finding the optimum wavelet scale at which to operate it. Following the example of other wavelet-based filters designed for point-source detection on CMB temperature maps \cite[e.g.,][]{amplificacion_patri}, a possible approach to optimum scale determination would be to identify the optimum filter scale as the one that maximizes the amplification, a quantity proportional to the quotient between the central wavelet coefficient amplitude and the wavelet coefficients dispersion. However, this maximum amplification criteria might not be the best approach for our characterization problem since it is one meant for \textit{detection}: it identifies the scale at which the source stands out more from the background, and there is no guarantee that the scale that maximizes amplification would be the one to minimize the error of the polarization angle estimate, an estimation that depends on a non-linear function like the arctangent. Although finding the optimum wavelet scale was outside the scope of the present work, it will be a topic of future study.\\

The ultimate goal of this project would be to apply the designed filter to real data. In particular, we are interested in applying the filter to the latest \emph{Planck} \texttt{NPIPE} data release \cite{npipe}, and comparing the polarization angle and intensity determined with standard methodologies like, for example, the one followed in the \textit{Second Planck Catalog of Compact Sources} \citep{PCCS_2015}, with the results our filter yields from an analysis in E- and B-modes. However, we will leave this endeavor to a future work since some possible sources of systematic error still have to be tested before applying the filter to real data. In particular, two of the possible sources of error that were bypassed in this work are the small distortions in the source profile introduced by the projection onto the plane, and the uncertainty in the location of the sources. For convenience, in our analysis we simulated point-sources directly on the plane instead of projecting them from the sphere. Although distortions are expected to be minimal, we still need to verify that the projection onto the plane does not introduce any additional bias to the polarization angle estimation. In addition, during our analysis we always assumed to know the exact position of the point-source in the patch. Although this seems to be a reasonable postulate since previous studies in intensity indicate that the uncertainty in the position of sources tends to lie below the pixel scale (see, for example, \cite{PCCS_2015}), during our study we noticed that polarization angle and intensity estimates can be very sensitive to the position and orientation of the source within the background, which suggests that the filter performance could also be significantly affected if it was not exactly applied on the center of the source. Intensity leakage and/or cross-polarization could also be another possible source of systematics. Following the example of more standard point-source detection methods, approaches like the one followed, for instance, in \cite{PCCS_2015}, could be implemented in order to correct for the contamination caused by band-pass mismatches.\\

Given the positive results obtained when studying extragalactic sources on E- and B-mode maps, the natural progression would be to extend the presented formalism from the characterization of known sources to an actual blind detection scheme operating on E- and B-mode maps. Catalogs made from E- and B-mode maps have the advantage of directly detecting polarization intensities, which makes no assumptions about the underlying polarization degree. In contrast, the usual strategy of detecting in intensity and subsequently looking at its counterpart on polarization tends to favor a certain range of polarization degrees, leaving out of the catalog sources with high polarization degrees that could have been detected in polarization despite not reaching the detection threshold in intensity. In this way, blind detection on E- and B-mode maps could help produce unbiased polarization degree catalogs.\\

\acknowledgments

PDP acknowledge partial financial support from the \emph{Formaci\'on del Profesorado Universitario (FPU) programme} of the Spanish Ministerio de Ciencia, Innovaci\'on y Universidades. DH acknowledges partial financial support from the Spanish Ministerio de Ciencia, Innovaci\'on y Universidades project PGC2018-101814-B-I00. The authors would like to thank the Spanish Agencia Estatal de Investigaci\'on (AEI, MICIU) for the financial support provided under the projects with references ESP2017-83921-C2-1-R and AYA2017-90675-REDC, co-funded with EU FEDER funds, and also acknowledge the funding from Unidad de Excelencia Mar\'ia de Maeztu (MDM-2017-0765). We make use of \verb|HEALPix| \cite{healpix}, and the \verb|numpy| and \verb|matplotlib| \cite{matplotlib} \verb|Python| packages.

\bibliographystyle{JHEP}
\bibliography{main}

\end{document}